\documentclass[pre,aps,preprint,preprintnumbers,amsmath,amssymb,secnumroman,eqsecnum]{revtex4}

\usepackage{amsmath}
\usepackage{amssymb}
\usepackage{graphicx}
\usepackage{dcolumn}
\usepackage{bm}

\newcommand{\beq}{\begin{equation}}
\newcommand{\eeq}{\end{equation}}
\newcommand{\beqa}{\begin{eqnarray}}
\newcommand{\eeqa}{\end{eqnarray}}
\newcommand{\vc}[1]{\mbox{\boldmath $#1$}}

\newcommand{\vol}[1]{{\bf #1}}

\begin{document}


\title{Optimizing second harmonic generation in a circular cylindrical waveguide with embedded periodically arranged tubelets of nonlinear susceptibility}

%

\author{B. U. Felderhof}

 \email{ufelder@physik.rwth-aachen.de}
\affiliation{Institut f\"ur Theorie der Statistischen Physik\\ RWTH Aachen University\\
Templergraben 55\\52056 Aachen\\ Germany\\
}%

\author{G. Marowsky}

 \email{gmarows@gwdg.de}
\affiliation{Laser-Laboratorium G\"ottingen e.V.\\
Hans-Adolf-Krebs-Weg 1\\37077 G\"ottingen\\ Germany\\
}%

\date{\today}

\begin{abstract}
Optical second-harmonic generation (SHG) is studied for the confined geometry of a circular cylindrical waveguide or optical fiber. A model situation of high symmetry is considered where the material with nonlinear susceptibility is isotropic and distributed in radially symmetric manner about the axis. In addition it is assumed that the material of high second-order nonlinearity consists of a thin circular layer, with periodic variation in the axial direction - similar to a usual quasi-phase-matched configuration. One can optimize the efficiency of SHG by choosing the period of the array such that a Bragg condition is satisfied. Depletion is studied in the framework of mode-coupling theory.
\end{abstract}

\maketitle

\section{\label{I}Introduction}

An efficient method of generating second-harmonic radiation (SHG) can find use in a variety of technical applications. Confined geometry - such as in fibers or film-waveguides - allows concentration of the fundamental radiation
at the position of the material with nonlinear susceptibility. In earlier work \cite{1} we have investigated SHG in a planar geometry, and studied the dependence on the position of nonlinear material with respect to the planar device that guides the fundamental radiation. SHG was enhanced by a judicious use of periodicity of the nonlinear material in the direction of propagation of the fundamental wave. In this paper we consider instead confinement of radiation in a circular waveguide or optical fiber. This geometry has the advantage of confinement in both transverse dimensions, thus avoiding diffraction and the corresponding dispersion of the fundamental beam in a transverse direction. The following calculations show that the circular geometry is preferable to the planar one, even when in the latter case the beam has infinite width, so that diffraction no longer plays a role.

In the following we analyze SHG in a circular waveguide for an idealized situation of high symmetry. The nonlinear material - representing an idealized polarization sheet \cite{1A}- is assumed to be arranged in a periodic array of
cylindrical tubelets centered around the axis of the waveguide. The period in the axial direction can be optimized by use of a Bragg condition involving the wavenumbers of both the fundamental and the doubled frequency. In addition one can optimize the axial width of the tubelets and their transverse radius. By integration over the radius the calculation can be extended to cover the case of tubelets of finite thickness, or of an array of solid cylindrical pieces. Admittedly, the idealized situation under consideration may be difficult to realize experimentally. The intention of our model calculation is to elucidate the principles, and to serve as a guide for the analysis of more realistic but less symmetric situations. Due to the high symmetry of the model situation we can limit attention to a small number of modes. This advantage is lost in more realistic situations.

The paper is organized as follows: First we perform a calculation in analogy to that for planar geometry \cite{1} for a finite length $L$ of nonlinear material distributed in $N$ tubelets, where $N$ is much larger than unity. For optimum phase matching the intensity of the generated SHG grows in proportion to $L^2$. In the circular waveguide the efficiency is sufficiently high so that it is necessary to consider depletion of the fundamental. On a large length scale the effect of depletion may be studied by use of mode-coupling theory. The situation is mathematically analogous to that of SHG in anisotropic crystals, so that the mode-coupling theory of Armstrong et al. \cite{2} can be used. Apparently this was not realized by Zhao et al. \cite{3}, who formulated mode-coupling equations on the much smaller length scale of the period of the array.

Second-harmonic generation in poled optical fibers using gratings optically written by mode interference was studied experimentally by Fermann et al. \cite {4}. Analogous experiments in thermally poled twin-hole glass fibers were performed by Mizunami et al. \cite{5},\cite{6}. Pump depletion in a waveguide filled with periodically poled lithium niobate was observed by Parameswaran et al. \cite{7} to be in good agreement with theory.

\section{\label{II}Circular waveguide theory}

We consider a circular waveguide of radius $b$ filled with material which is uniform in the axial direction $z$ and with electric and magnetic permeability, which depend only on the radial direction $r$. We use cylindrical coordinates $(r,\varphi,z)$. The dielectric profile $\varepsilon(r,\omega)$ and magnetic profile $\mu(r,\omega)$ depend also on frequency $\omega$. We assume that for the frequencies of interest $\varepsilon$ and $\mu$ are real. Also we assume that $\varepsilon$ and $\mu$ tend to constants $\varepsilon_i,\mu_i$ for small $r$ and to constants $\varepsilon_f,\mu_f$ for $r\rightarrow b$. In later application we consider in particular a two-layer situation with $\varepsilon,\mu=\varepsilon_2,\mu_2$ for $0<r<d$ and $\varepsilon,\mu=\varepsilon_1,\mu_1$ for $d<r<b$.

We consider plane wave solutions of Maxwell's equations which depend on $z$ and $t$ through a factor $\exp(ipz-i\omega t)$, and which do not depend on the azimuthal angle $\varphi$. Maxwell's equations for the electric and magnetic field amplitudes then read in Gaussian units \cite{8}-\cite{10}
\begin{eqnarray}
\label{2.1}\frac{d\varepsilon rE_r}{dr}+ip\varepsilon rE_z=0,\qquad\frac{d\mu rH_r}{dr}+ip\mu rH_z=0, \nonumber\\
\frac{dE_\varphi}{dr}+\frac{E_\varphi}{r}=ik\mu H_z,\qquad\frac{dH_\varphi}{dr}+\frac{H_\varphi}{r}=-ik\varepsilon E_z,\nonumber\\
\frac{dE_z}{dr}-ipE_r=-ik\mu H_\varphi,\qquad\frac{dH_z}{dr}-ipH_r=ik\varepsilon E_\varphi,\nonumber\\
pE_\varphi=-k\mu H_r,\qquad pH_\varphi=k\varepsilon E_r,
\end{eqnarray}
where $k=\omega/c$ with velocity of light $c$ is the vacuum wavenumber. The solutions of these equations may be decomposed according to two polarizations. For TE-polarization the components $E_r,\;E_z$, and $H_\varphi$ vanish, and the equations may be combined into the single equation
\begin{equation}
\label{2.2}\frac{d^2E_\varphi}{dr^2}-\frac{r}{\mu}\frac{d(\mu/r)}{dr}\frac{dE_\varphi}{dr}+\bigg(\varepsilon\mu k^2-\frac{1}{\mu r}\frac{d\mu}{dr}-\frac{1}{r^2}\bigg) E_\varphi=p^2E_\varphi\qquad (\mathrm{TE}).
\end{equation}
For TM-polarization the components $H_r,\;H_z$, and $E_\varphi$ vanish, and the equations may be combined into the single equation
\begin{equation}
\label{2.3}\frac{d^2H_\varphi}{dr^2}-\frac{r}{\varepsilon}\frac{d(\varepsilon/r)}{dr}\frac{dH_\varphi}{dr}+\bigg(\varepsilon\mu k^2-\frac{1}{\varepsilon r}\frac{d\varepsilon}{dr}-\frac{1}{r^2}\bigg) H_\varphi=p^2H_\varphi\qquad (\mathrm{TM}).
\end{equation}

We consider first TE-polarization. We assume that for $0<r<r_i$ the permeabilities $\varepsilon,\mu$ equal $\varepsilon_i,\mu_i$ , and that for $r>r_f$ they equal $\varepsilon_f,\mu_f$. We write the solution of Eq. (2.2) in these two regions
 \begin{eqnarray}
\label{2.4}E_\varphi(r)=J_1(q_ir)\qquad\mathrm{for}\;r<r_i,\nonumber\\
E_\varphi(r)=A_fJ_1(q_fr)+B_fY_1(q_fr)\qquad\mathrm{for}\;r>r_f,
\end{eqnarray}
with Bessel-functions $J_1(qr),\;Y_1(qr)$ and wavenumbers
\begin{equation}
\label{2.5}q_j=\sqrt{\varepsilon_j\mu_jk^2-p^2}.
\end{equation}
The coefficients $A_f$ and $B_f$ are related by the boundary conditions at $r=b$. The wavenumbers $\{q_i,q_f\}$ are real only up to a maximum value of $p$ given by $\sqrt{\varepsilon_j\mu_j}k$ in either case, and are pure imaginary beyond this value. For such wavenumbers we rewrite the second equation in (2.4) as
\begin{equation}
\label{2.6}E_\varphi(r)=AI_1(\kappa_f r)+BK_1(\kappa_f r)\qquad\mathrm{for}\;r>r_f,
\end{equation}
with modified Bessel-functions $I_1(\kappa r)$ and $K_1(\kappa r)$ and $\kappa=\sqrt{p^2-\varepsilon\mu k^2}$. The guided mode solutions occur at discrete values $\{p_j\}$ of $p$ larger than $\sqrt{\varepsilon_f\mu_f}k$.  We assume that the radius $b$ is sufficiently large that the waveguide condition can be approximated by
\begin{equation}
\label{2.7}A(p,k)=0.
\end{equation}
At fixed $k$ the roots of this equation determine the discrete values $\{p_j\}$ for which a guided mode solution exists. We call $A(p,k)$ the dispersion function.

The one-dimensional wave equation (2.2) can be transformed to a form resembling the one-dimensional time-independent Schr\"odinger equation by use of the transformation
\begin{equation}
\label{2.8}E_\varphi(r)=\sqrt{\frac{\mu}{r}}f(r).
\end{equation}
By substitution we find that Eq. (2.2) is transformed to
\begin{equation}
\label{2.9}\frac{d^2f}{dr^2}-V(r)f=p^2f,
\end{equation}
 where the function $V(r)$ is given by
 \begin{equation}
\label{2.10}V(r)=-\varepsilon\mu k^2+\sqrt{\frac{\mu}{r}}\frac{d^2}{dr^2}\sqrt{\frac{r}{\mu}}+\frac{1}{r}\frac{d\log(\mu r)}{dr}.
\end{equation}

 It is of interest to derive an expression for the norm of the eigensolutions. Differentiating Eq. (2.9) with respect to $p^2$ one derives the identity
 \begin{equation}
\label{2.11}\frac{\partial}{\partial r}\bigg[f\frac{\partial}{\partial r}\bigg(\frac{\partial f}{\partial p^2}\bigg)-\frac{\partial f}{\partial p^2}\frac{\partial f}{\partial r}\bigg]=f^2.
\end{equation}
Applying this identity to guided mode solutions normalized as in Eq. (2.4) we find by integration over $r$ and use of the waveguide condition (2.7)
 \begin{equation}
\label{2.12}N^E_j=\int^b_0\frac{r}{\mu(r)}E_{\varphi j}^2(r)\;dr=\frac{1}{2\mu_fp_j}B(p_j,k)\frac{\partial A(p,k)}{\partial p}\bigg|_{p_j}.
\end{equation}
We shall show in the next section that the norm is related to the intensity of the mode.

It follows from Eq. (2.3) that for TM-polarization exactly the same relations hold if we replace $E_\varphi$ by $H_\varphi$, $\varepsilon$ by $\mu$,
and $\mu$ by $\varepsilon$. Where necessary the symbols corresponding to the two types of solution will be distinguished by a superscript E or M.

\section{\label{III}Excitation of guided modes}

In this section we describe how the eigenmodes may be excited by an oscillating dipole density. We begin by relating the norm of an eigenmode, given by Eq. (2.12), to the physical intensity. The energy current density averaged over a time period $2\pi/\omega$ is given by the Poynting vector
\begin{equation}
\label{3.1}\vc{S}=\frac{c}{8\pi}\mathrm{Re}(\vc{E}\times\vc{H}^*).
\end{equation}
The $z$-component of this expression may be decomposed into
\begin{equation}
\label{3.2}S_z=S_z^E+S_z^M,
\end{equation}
with the separate terms for TE- and TM-polarization
\begin{equation}
\label{3.3}S_z^E=\frac{c}{8\pi}\frac{p}{k\mu}|E_\varphi|^2,\qquad S_z^M=\frac{c}{8\pi}\frac{p}{k\varepsilon}|H_\varphi|^2.
\end{equation}
We define the total intensity by the integrals
\begin{equation}
\label{3.4}I^E=2\pi\int^b_0S^E_z(r)r\;dr,\qquad I^M=2\pi\int^b_0S^M_z(r)r\;dr.
\end{equation}
The intensity does not depend on $z$.

For a single eigenmode of either TE- or TM-type the expressions (3.3) become
\begin{equation}
\label{3.5}S_{zj}^E=\frac{c}{8\pi}\frac{p_j^E}{k\mu}|E_{\varphi j}|^2,\qquad S_{zj}^M=\frac{c}{8\pi}\frac{p_j^M}{k\varepsilon}|H_{\varphi j}|^2.
\end{equation}
By comparison with Eq. (2.12) we find that for a single eigenmode excited with amplitude $a_j$ the intensity is related to the norm of the mode by
\begin{equation}
\label{3.6}I_j=\frac{cp_j}{4k}N_j|a_j|^2.
\end{equation}
This expression is formally the same for both polarizations.

Using orthogonality of the eigenmodes \cite{8}-\cite{10} we find that for a linear superposition of guided modes, all oscillating at the same frequency $\omega$, the total intensity is given by
\begin{equation}
\label{3.7}I=\sum_j(I^E_j+I^M_j),
\end{equation}
which again does not depend on $z$.

Next we investigate how radiation is emitted by an antenna embedded in the waveguide. We consider a surface polarization \cite{11},\cite{12} $\vc{P}^S(z)$ of finite extent in the $z$-direction, independent of $\varphi$, located at radius $r=r_0$, and oscillating at frequency $\omega$. The corresponding charge and current densities are
 \begin{eqnarray}
\label{3.8}\rho(\vc{r})=-\frac{\partial P^S_z}{\partial z}\delta(r-r_0)-P^S_z\delta'(r-r_0),\nonumber\\
\vc{j}(\vc{r})=-i\omega\vc{P}^S(z)\delta(r-r_0),
\end{eqnarray}
which must be added as source terms to Maxwell's equations. A Fourier analysis of the surface polarization yields
\begin{equation}
\label{3.9}\vc{P}^S(z)=\int\hat{\vc{P}}^S(p)e^{ipz}\;dp.
\end{equation}
From Maxwell's equations we now find instead of Eq. (2.2)
\begin{eqnarray}
\label{3.10}\frac{d^2\hat{E}_\varphi}{dr^2}-\frac{r}{\mu}\frac{d(\mu/r)}{dr}\frac{d\hat{E}_\varphi}{dr}+\bigg(\varepsilon\mu k^2-p^2-\frac{1}{\mu r}\frac{d\mu}{dr}-\frac{1}{r^2}\bigg) \hat{E}_\varphi\nonumber\\=
-4\pi k^2\mu\hat{P}^S_\varphi\delta(r-r_0).
\end{eqnarray}
Similarly instead of Eq. (2.3)
\begin{eqnarray}
\label{3.11}\frac{d^2\hat{H}_\varphi}{dr^2}-\frac{r}{\varepsilon}\frac{(d\varepsilon/r)}{dr}\frac{d\hat{H}_\varphi}{dr}+\bigg(\varepsilon\mu k^2-p^2-\frac{1}{\varepsilon r}\frac{d\varepsilon}{dr}-\frac{1}{r^2}\bigg) \hat{H}_\varphi=\nonumber\\-4\pi ik\bigg[\hat{P}^S_z\delta'(r-r_0)-\frac{1}{\varepsilon}\frac{d\varepsilon}{dr}\hat{P}^S_z\delta(r-r_0)\bigg]-4\pi kp\hat{P}^S_r\delta(r-r_0).
\end{eqnarray}

The solution of these equations may be found with the aid of the Green's functions $G^E(r,r_0)$ and $G^M(r,r_0)$ defined by the equation
\begin{equation}
\label{3.12}\frac{d^2G}{dr^2}-V(r)G=p^2G+\delta(r-r_0).
\end{equation}
The Green's function may be expressed as
\begin{equation}
\label{3.13}G(r,r_0)=\frac{f_1(r_<)f_2(r_>)}{\Delta(f_1,f_2)},
\end{equation}
where $r_<(r_>)$ is the smaller (larger) of $r$ and $r_0$, in terms of the two fundamental solutions $f_1,\;f_2$ defined for $p>\sqrt{\varepsilon_f\mu_f}k$ by
\begin{eqnarray}
\label{3.14}f^E_1(r)&=&\sqrt{\frac{r}{\mu_i}}\;J_1(q_ir),\qquad f^M_1(r)=\sqrt{\frac{r}{\varepsilon_i}}\;J_1(q_ir),\qquad\mathrm{for}\;r<r_i,\nonumber\\
f^E_2(r)&=&\sqrt{\frac{r}{\mu_f}}\;K_1(\kappa_fr),\qquad f^M_2(r)=\sqrt{\frac{r}{\varepsilon_f}}\;K_1(\kappa_fr),\qquad\mathrm{for}\;r>r_f,
\end{eqnarray}
with the Wronskian
\begin{equation}
\label{3.15}\Delta(f_1,f_2)=\bigg|\begin{array}{cc}f_1&f_2\\f'_1&f'_2\end{array}\bigg|.
\end{equation}
The Wronskian takes the value
\begin{equation}
\label{3.16}\Delta(f^E_1,f^E_2)=\frac{-1}{\mu_f}A^E(p,k),\qquad\Delta(f^M_1,f^M_2)=\frac{-1}{\varepsilon_f}A^M(p,k).
\end{equation}

The solution of Eq. (3.10) is given by
\begin{equation}
\label{3.17}\hat{E}_\varphi(p,r)=-4\pi k^2[\mu(r)/r]^{1/2}G^E(r,r_0)[\mu(r_0)r_0]^{1/2}\hat{P}^S_\varphi.
\end{equation}
The solution of Eq. (3.11) is given by
\begin{eqnarray}
\label{3.18}\hat{H}_\varphi(p,r)=4\pi ik[\varepsilon(r)/r]^{1/2}\bigg[\frac{\partial G^M(r,r_0)}{\partial r_0}+\frac{1}{2}\frac{d\ln\varepsilon(r_0)}{dr_0}G^M(r,r_0)\nonumber\\
+\frac{1}{2r_0}G^M(r,r_0)\bigg][\varepsilon(r_0)/r_0]^{-1/2}\hat{P}^S_z
-4\pi kp[\varepsilon(r)/r]^{1/2}G^M(r,r_0)[\varepsilon(r_0)/r_0]^{-1/2}\hat{P}^S_r.
\end{eqnarray}

For a surface polarization $\vc{P}^S(z)$ with arbitrary variation in the $z$-direction the fields $E_\varphi(r,z)$ and $H_\varphi(r,z)$ are now obtained by Fourier superposition. Thus we find
\begin{equation}
\label{3.19}E_\varphi(r,z)=\int \hat{E}_\varphi(p,r)e^{ipz}\;dp,\qquad
H_\varphi(r,z)=\int \hat{H}_\varphi(p,r)e^{ipz}\;dp,
\end{equation}
where $\hat{E}_\varphi(p,r)$ and $\hat{H}_\varphi(p,r)$ are given by Eqs. (3.17) and (3.18) in terms of $\hat{\vc{P}}^S(p)$. The Wronskian, given by Eq. (3.16) vanishes at the eigenvalues $\{p^E_j\}$ and $\{p^M_j\}$. For large $z$ the contribution from the corresponding poles dominates the integrals in Eq. (3.19). This allows us to evaluate the amplitude of the various guided modes excited by the oscillating surface polarization.

\section{\label{IV}Emitted radiation}

In this section we analyze the radiation emitted by a circular cylindrical antenna, as introduced in the preceding section, in more detail. We are interested in the radiation channeled into the waveguide and detected at large positive $z$. At sufficiently large distance from the antenna, i.e. after the decay of transients corresponding to evanescent wave solutions, the behavior of the fields is dominated by pole contributions to the integrals, corresponding to roots of the waveguide condition (2.7). The contributions may be found by contour integration in Eq. (3.19) with the poles at positive $\{p_j\}$ shifted slightly upwards into the complex plane, and those at $\{-p_j\}$ shifted slightly downwards. We note that it follows from Eq. (2.6) that for values $\{p_j\}$ for which the waveguide condition is satisfied
\begin{equation}
\label{4.1}f_1(p_j,r)=B(p_j,k)f_2(p_j,r).
\end{equation}
Thus we find for large positive $z$
\begin{equation}
\label{4.2}E_\varphi(r,z)\approx\sum_ja_{1j}^E\psi^E_j(r)\exp(ip^E_jz),
\end{equation}
where we employ the notation
\begin{equation}
\label{4.3}\psi^E_j(r)=\sqrt{\mu(r)/r}f_{1j}^E(r).
\end{equation}
The amplitudes are given by
\begin{equation}
\label{4.4}a_{1j}^E=4\pi^2ik^2(p^E_jN^E_j)^{-1}r_0\psi^E_j(r_0)\hat{P}^S_\varphi(p^E_j),
\end{equation}
where we have used Eq. (2.12). From Eq. (4.2) we may evaluate the intensity defined in Eq. (3.4). Because of the orthogonality of the different modes \cite{8}-\cite{10} there are no cross terms, and we find by use of Eq. (3.6)
\begin{equation}
\label{4.5}I^E=\sum_jI^E_j=4\pi^4ck^3\sum_j(p^E_jN^E_j)^{-1}r_0^2[\psi^E_j(r_0)]^2\;|\hat{P}^S_\varphi(p^E_j)|^2.
\end{equation}
Similarly we find for large positive $z$
\begin{equation}
\label{4.6}H_\varphi(r,z)\approx\sum_ja_{1j}^M\psi^M_j(r)\exp(ip^M_jz),
\end{equation}
with the notation
\begin{equation}
\label{4.7}\psi^M_j(r)=\sqrt{\varepsilon(r)/r}f_{1j}^M(r).
\end{equation}
The amplitudes are given by
\begin{equation}
\label{4.8}a_{1j}^M=4\pi^2k(p^M_jN^M_j)^{-1}\big[R_j(r_0)\hat{P}^S_r(p^M_j)+Z_j(r_0)\hat{P}^S_z(p^M_j)
\big],
\end{equation}
with the abbreviations
\begin{eqnarray}
\label{4.9}
R_j(r)&=&\frac{ir}{\varepsilon(r)}p^M_j\psi^M_j(r),\nonumber\\
Z_j(r)&=&\frac{r}{\varepsilon(r)}\bigg(\frac{\partial\psi^M_j(r)}{\partial r}+\frac{1}{r}\psi^M_j(r)\bigg).
\end{eqnarray}
This yields for the intensity
\begin{eqnarray}
\label{4.10}I^M&=&\sum_jI^M_j\nonumber\\
&=&4\pi^4ck\sum_j(p^M_jN^M_j)^{-1}\big|R_j(r_0)\hat{P}^S_r(p^M_j)+Z_j(r_0)\hat{P}^S_z(p^M_j)
\big|^2.
\end{eqnarray}
The expressions (4.5) and (4.10) have a fairly simple structure. The efficiency with which a surface polarization $\vc{P}^S(z)$ excites the guided modes is determined by its Fourier component at the wavenumber $p_j$, as well as by its radial location $r_0$ via the eigenfunction $\psi_j(r_0)$, which appears quadratically with its proper normalization $N_j$.

\section{\label{V}Phase-matching}

In this section we discuss the principle of second harmonic generation by use of a phase-matched adsorbate embedded in a circular cylindrical waveguide. We shall assume that the adsorbate either is located as a thin layer directly outside the core, or is embedded in the core. We consider a surface polarization $\vc{P}^S(z)$ located at radius $r_0$ and induced by an incident fundamental wave. The polarization acts as an antenna emitting waves at the second harmonic frequency. Thus we put
\begin{equation}
\label{5.1}\vc{P}^S(r_0,z)=\vc{\chi}^{(2)}(z):\vc{E}_{10}(r_0,z)\vc{E}_{10}(r_0,z),
\end{equation}
where $\vc{E}_{10}(r_0,z)$ is the incident fundamental field at the location of the adsorbate. If the fundamental field oscillates at frequency $\omega$, then the surface polarization oscillates at frequency $2\omega$, and this must be taken into account in the expressions of the preceding sections. We assume that the adsorbate is so weak that it does not disturb the fundamental wave. This is expressed by the subscript zero in (5.1).

The fundamental wave is a linear combination of guided modes with $z$-dependence $\exp(ip_jz)$ with wavenumber $p_j(\omega)$. We shall assume that the susceptibility $\vc{\chi}^{(2)}$ depends on $z$ via the density of adsorbed molecules. If the adsorbate has a periodicity in the $z$- direction with period $a$ characterized by the wavenumber $K=2\pi/a$, then we may expect resonance when the phase-matching condition
\begin{equation}
\label{5.2}p_j(2\omega)=2p_k(\omega)+nK,\qquad n=0,\pm1,\pm2,...
\end{equation}
is satisfied. More specifically it is natural to aim at satisfying the condition
\begin{equation}
\label{5.3}p_0(2\omega)=2p_0(\omega)\pm K,
\end{equation}
for the lowest mode $j=0$. We shall call this the Bragg condition.

We consider in particular a grating of period $a$ consisting of $N$ adsorbate tubelets of width $w<a$. An example of the grating and waveguide is shown in Figs. 1 and 2. The susceptibility function is given by
\begin{equation}
\label{5.4}\vc{\chi}(z)=\vc{\chi}^{(2)} g_N(z),
\end{equation}
with the Bragg function
\begin{equation}
\label{5.5}g_N(z)=\sum^{N-1}_{n=0}\theta(w,z-na-\frac{1}{2}w),
\end{equation}
where
\begin{eqnarray}
\label{5.6}\theta(w,z)=1\qquad\mathrm{for}\;-\frac{w}{2}<z<\frac{w}{2},\nonumber\\
=0\qquad\mathrm{for}\;|z|>\frac{w}{2}.
\end{eqnarray}
The prefactor $\vc{\chi}^{(2)}$ in (5.4) is a third rank tensor independent of $z$. We shall assume that $\vc{\chi}^{(2)}$ has the characteristics  of a layer isotropic about the surface normal in the radial direction. We also assume that there exists a mirror plane containing the radial normal to exclude chirality. From these assumptions it follows that $\vc{\chi}^{(2)}$ has only three independent components (for details see \cite{13}-\cite{16}). The Fourier-component $\hat{\vc{P}}^S(p)$ is proportional to
\begin{eqnarray}
\label{5.7}\hat{G}_N(p-2p_k(\omega))&=&\frac{1}{2\pi}\int^\infty_{-\infty} g_N(z)e^{2ip_k(\omega)z-ipz}\;dz\nonumber\\
&=&\frac{\exp(is_kw)-1}{2\pi is_k}\;\frac{1-\exp{(iNs_ka)}}{1-\exp(is_ka)},
\end{eqnarray}
where we have introduced the variable
\begin{equation}
\label{5.8}s_k=2p_k(\omega)-p.
\end{equation}
For real $p$ the absolute square of the second factor in Eq. (5.7) is given by
 \begin{equation}
\label{5.9}F_N(sa)=\bigg|\frac{1-e^{iNsa}}{1-e^{isa}}\bigg|^2=\frac{\sin^2(Nsa/2)}{\sin^2(sa/2)},
\end{equation}
which takes the values $N^2$ at $sa=2n\pi$, where $n=0,\pm1,\pm2,...$. Since we wish $\hat{\vc{P}}^S(p)$ to be maximum at $p_0(2\omega)$ we choose the lattice distance $a$ such that
 \begin{equation}
\label{5.10}a=2\pi|2p_0(\omega)-p_0(2\omega)|^{-1}
\end{equation}
corresponding to $n=1$ or $n=-1$. In this way we satisfy the Bragg condition (5.3). The width of the function in Eq. (5.9) at $s=\pm2\pi/a$ is of order $1/Na$. Hence the area of the peak is proportional to $N$. The larger $N$, the more precisely the condition (5.10) must be satisfied. An error $\Delta a$ in the value of $a$ implies an error $\Delta p=2\pi\Delta a/a^2$ in $p$-space. If we require this to be at most $1/Na$, then $N$ cannot be larger than $a/2\pi\Delta a$. Ideally one would use a tunable laser and adjust the frequency $\omega$ such that the condition (5.10)  is precisely satisfied. With that choice the absolute square of the first factor in Eq. (5.7) is maximal at $w=\frac{1}{2}a$.

\section{\label{VI}Second harmonic generation}

In this section we investigate the effect of geometry on the efficiency for second harmonic generation. We consider the core of the waveguide to be a cylinder of radius $d$ with dielectric constant $\varepsilon_2$ surrounded by an outer mantle with uniform dielectric constant $\varepsilon_1<\varepsilon_2$. The grating of adsorbed molecules is located at radius $r_0$, either inside or outside the core ($r_0<d$ or $r_0>d$). We put the magnetic permeability equal to unity everywhere, and consider guided mode solutions of TM-type. We study the dependence of the efficiency for second harmonic generation on the radius $r_0$.

The discontinuity of the dielectric constant at $r=d$ corresponds to jump conditions for the tangential component $H_\varphi(r)$. The first condition is that $H_\varphi(r)$ is continuous at $r=d$. The wave equation (2.3) may be rewritten as
 \begin{equation}
\label{6.1}\varepsilon(r)\frac{d}{dr}\bigg[\frac{1}{\varepsilon(r)}\frac{dH_\varphi}{dr}+\frac{H_\varphi}{r\varepsilon(r)}\bigg]+\varepsilon\mu k^2H_\varphi=p^2H_\varphi.
\end{equation}
Hence the second condition is that $(dH_\varphi/dr+H_\varphi/r)/\varepsilon$ is continuous at the interface. In analogy to Eqs. (2.4) and (2.6) we write the solution as
\begin{eqnarray}
\label{6.2}H_\varphi(r)&=&J_1(q_2r),\qquad\mathrm{for}\;0<r<d,\nonumber\\
&=&A^MI_1(\kappa_1r)+B^MK_1(\kappa_1r),\qquad\mathrm{for}\;d<r<b,
\end{eqnarray}
where $q_2=\sqrt{\varepsilon_2k^2-p^2}$ and $\kappa_1=\sqrt{p^2-\varepsilon_1k^2}$. From the two continuity equations we find for the coefficients $A^M$ and $B^M$
\begin{eqnarray}
\label{6.3}A^M(p,k)&=&\frac{\varepsilon_1}{\varepsilon_2}\;q_2dJ_0(q_2d)K_1(\kappa_1d)+\kappa_1dJ_1(q_2d)K_0(\kappa_1d),\nonumber\\
B^M(p,k)&=&-\frac{\varepsilon_1}{\varepsilon_2}\;q_2dJ_0(q_2d)I_1(\kappa_1d)+\kappa_1dJ_1(q_2d)I_0(\kappa_1d).
\end{eqnarray}
Putting $A^M(p,k)=0$ for fixed $k$ one finds the wavenumbers $p^M_j(k)$ of the guided modes. The guided mode solutions take the form
\begin{eqnarray}
\label{6.4}\psi^M_j(r)&=&J_1(q_{2j}r),\qquad\mathrm{for}\;0<r<d,\nonumber\\
&=&B^M(p^M_j,k)K_1(\kappa_{1j}r),\qquad\mathrm{for}\;d<r<b.
\end{eqnarray}
From Eq. (2.12) one finds for their norm
 \begin{equation}
\label{6.5}N^M_j=\int^b_0\frac{r}{\varepsilon(r)}[\psi^M_J(r)]^2\;dr=\frac{1}{2\varepsilon_1p^M_j}B^M(p_j,k)\frac{\partial A^M(p,k)}{\partial p}\bigg|_{p^M_j}.
\end{equation}

The electrical field has components
\begin{equation}
\label{6.6}E_r(r)=\frac{p}{k\varepsilon}H_\varphi,\qquad
E_z(r)=\frac{i}{k\varepsilon}\bigg(\frac{dH_\varphi}{dr}+\frac{H_\varphi}{r}\bigg).
\end{equation}
We assume that the fundamental is present as a single mode oscillating at frequency $\omega$ with amplitude $a^M_{1k}(\omega)$. Hence the electrical field vector is
\begin{equation}
\label{6.7}\vc{E}(\omega;r,z)=\frac{a^M_{1k}(\omega)}{k\varepsilon(r)}\bigg[p^M_k\psi^M_k(r)\vc{e}_r
+i\bigg(\frac{d\psi^M_k(r)}{dr}+\frac{\psi^M_k(r)}{r}\bigg)\vc{e}_z\bigg]\exp(ip^M_kz).
\end{equation}
We assume an isotropic tensor $\vc{\chi}^{(2)}$ with mirror symmetry. The induced surface polarization is
\begin{equation}
\label{6.8}\vc{P}^{SM}(2\omega,z)=g_N(z)\bigg(\frac{ca^M_{1k}(\omega)}{\omega\varepsilon(\omega,r_0)}\bigg)^2\vc{X}_k(\omega,r_0)\exp(2ip^M_kz),
\end{equation}
with Bragg factor given by Eq. (5.5) and vector $\vc{X}_k(\omega,r_0)$ given by
\begin{eqnarray}
\label{6.9}\vc{X}_k(\omega,r)&=&\bigg[\chi_1\big(p^M_k\psi^M_k(\omega,r)\big)^2
-\chi_2\bigg(\frac{d\psi^M_k(r)}{dr}+\frac{\psi^M_k(r)}{r}\bigg)^2\bigg]\vc{e}_r\nonumber\\
&+&2i\chi_3p^M_k\psi^M_k(\omega,r)\bigg(\frac{d\psi^M_k(r)}{dr}+\frac{\psi^M_k(r)}{r}\bigg)\vc{e}_z,
\end{eqnarray}
where $\chi_1=\chi^{(2)}_{rrr},\;\chi_2=\chi^{(2)}_{rzz}$ and $\chi_3=\chi^{(2)}_{zrz}=\chi^{(2)}_{zzr}$ are the relevant components of the nonlinear susceptibility tensor $\vc{\chi}^{(2)}$. We assume that the layer of second order susceptibility is locally flat, so that the subscript $r$ corresponds  to the locally normal component, and $z$ to the locally tangential component. For a representation of the tensor in the local Cartesian frame see Roders et al. \cite{16}. The emitted second harmonic radiation is TM-polarized. The intensity of the emitted second harmonic radiation is given by Eq. (4.10), with the right-hand side taken at frequency $2\omega$ instead of $\omega$. We find by use of Eq. (3.6)
\begin{eqnarray}
\label{6.10}I^M_j(2\omega)&=&\frac{128\pi^4}{\omega\varepsilon(\omega,r_0)^4}|\hat{G}_N(p^M_j(2\omega)-2p^M_k(\omega))|^2\frac{1}{p^M_j(2\omega)N^M_j(2\omega)p^M_k(\omega)^2N^M_k(\omega)^2}\nonumber\\
&\times&\big|R_j(2\omega,r_0)X_{kr}(\omega,r_0)+Z_j(2\omega,r_0)X_{kz}(\omega,r_0)\big|^2\big(I^M_k(\omega)\big)^2.
\end{eqnarray}
The conversion coefficient is defined by
\begin{equation}
\label{6.11}\eta^{MM}_{jk}=I^M_j(2\omega)/I^M_k(\omega).
\end{equation}

We write the conversion coefficient in the form
\begin{equation}
\label{6.12}\eta^{MM}_{jk}=\frac{128\pi^4}{c}\;|\hat{G}_N(p^M_j(2\omega)-2p^M_k(\omega))|^2A_{jk}(r_0)I^M_k(\omega),
\end{equation}
with coefficient
\begin{equation}
\label{6.13}A_{jk}(r_0)=\frac{\big|R_j(2\omega,r_0)X_{kr}(\omega,r_0)+Z_j(2\omega,r_0)X_{kz}(\omega,r_0)\big|^2}{k\varepsilon(\omega,r_0)^4p^M_j(2\omega)N^M_j(2\omega)p^M_k(\omega)^2N^M_k(\omega)^2},
\end{equation}
and write the latter as
\begin{equation}
\label{6.14}A_{jk}(r_0)=d^{-6}\sum_{\lambda\mu}C_{\lambda\mu}(j|k;r_0)\chi_\lambda\chi^*_\mu,
\end{equation}
with dimensionless coefficients $C_{\lambda\mu}(j|k;r_0)$. The factors $R_j$ and $Z_j$ are given by Eq. (4.9), and $X_r$ and $X_z$ are given by Eq. (6.9). The coupling coefficients $C_{\lambda\mu}(j|k)$ depend on the radius $r_0$ and the geometry of the waveguide. The efficiency of second harmonic generation for the chosen geometry is characterized by the Bragg prefactor $|\hat{G}_N|^2$ and the coefficients $C_{\lambda\mu}(j|k;r_0)$.

We consider the width $d$ to be fixed, and vary the frequency $\omega$. The input laser is tuned in a narrow frequency range, so that dispersion of the dielectric constant near $\omega$ and $2\omega$ may be neglected. We put $\varepsilon_2(\omega)=2.25$, $\varepsilon_2(2\omega)=2.28$, in combination with $\varepsilon_1(\omega)=2.13$ and $\varepsilon_1(2\omega)=2.17$.
In Fig. 3 we plot the reduced wavenumbers $p_0(k)/k, p_1(k)/k$ and $p_2(k)/k$ at the fundamental frequency as functions of $kd$. The ratios $\{p_n(k)/k\}$ are larger than $\sqrt{\varepsilon_1}=1.4595$ and less than $\sqrt{\varepsilon_2}=1.5$. The corresponding plots at the second harmonic frequency are very similar.

In Fig. 4 we plot the coefficient $C_{11}(0|0;r_0)$ at $kd$=24 as a function of
the fraction $r_0/d$. In Figs. 5 and 6 we present similar plots for the coefficients $C_{22}(0|0;r_0)$ and $C_{33}(0|0;r_0)$. The plot in Fig. 4 for susceptibility $\chi_1=\chi^{(2)}_{rrr}$ shows the largest rate of conversion.

It is of interest to compare the optimal situation with that for a planar waveguide. We have found earlier that for a planar waveguide consisting of a slab of dielectric constant $\varepsilon_2$ of thickness $2d$ surrounded by a medium of dielectric constant $\varepsilon_1$ second harmonic generation is optimal for a thin polarization layer midway between the two interfaces (in Ref. 1 this was called geometry III). For the planar waveguide the conversion coefficient took the form
\begin{equation}
\label{6.15}\eta^{MM}_{jkP}=\frac{256\pi^5}{\omega}\;G_{NjkP}\bigg(\sum_{\lambda\mu}B^{\lambda\mu}_{jk}\chi_\lambda\chi^*_\mu\bigg)J^M_k(\omega),
\end{equation}
with Bragg factor
\begin{equation}
\label{6.16}G_{NjkP}=|\hat{G}_N(p^M_j(2\omega)-2p^M_k(\omega))|^2.
\end{equation}
Here the wavenumbers $p$ must be calculated for the guided modes of the planar waveguide, so that the lattice distance $a$ and the Bragg factor $G_{NjkP}$ differ from the lattice distance and corresponding factor $G_{NjkC}$ for a cylindrical waveguide appearing in Eq. (6.12). The intensity $J^M_k(\omega)$ is defined from the integral of the Poynting vector along the transverse coordinate. In order to compare the two geometries we put
\begin{equation}
\label{6.17}I^M_k(\omega)=\pi d^2\overline{S}_C,\qquad J^M_k(\omega)=2d\overline{S}_P,
\end{equation}
and write the conversion coefficient (6.12) for the cylindrical waveguide in the form
\begin{equation}
\label{6.18}\eta^{MM}_{jkC}=128\pi^5G_{NjkC}A_{jkC}(r_0)\frac{d^2\overline{S}_C}{c},
\end{equation}
where $G_{NjkC}$ is the Bragg factor defined as in Eq. (6.16). The conversion coefficient for the planar waveguide is written similarly as
\begin{equation}
\label{6.19}\eta^{MM}_{jkP}=128\pi^5G_{NjkP}A_{jkP}\frac{d^2\overline{S}_P}{c}.
\end{equation}
By comparison with Eq. (6.15)
\begin{equation}
\label{6.20}A_{jkP}=\frac{4}{kd}\sum_{\lambda\mu}B^{\lambda\mu}_{jk}\chi_\lambda\chi^*_\mu.
\end{equation}

We consider $kd=24$, as in the experiment of Parameswaran et al. \cite{7}, with wavelength of the fundamental  $\lambda=2\pi/k=1550\;\mathrm{nm}$ and $d\approx 6\mu\mathrm{m}$, and use the same dielectric constants as above. Then we find for the cylindrical waveguide $p_0(\omega)d=35.837$ and $p_0(2\omega)d=72.388$, and for the planar waveguide $p_0(\omega)d=35.972$ and $p_0(2\omega)d=72.463$. The lattice distance $a$, given by Eq. (5.10) in the two cases takes the value
\begin{equation}
\label{6.21}a_C=8.79\;d,\qquad a_P=12.12\;d.
\end{equation}
Correspondingly we find for the coupling factors in Eqs. (6.18) and (6.19) for $\chi_2=\chi_3=0$
\begin{equation}
\label{6.22}A_{00C}(r_m)=4275\;|\chi_1|^2/d^6,\qquad A_{00P}=32.47\;|\chi_1|^2/d^6.
\end{equation}
for the value $r_m=0.593\;d$ corresponding to the maximum in Fig. 4. The prefactors in Eq. (6.22) are the values of the coupling coefficients $C_{11C}(0|0)=4275$ and $C_{11P}(0|0)=32.47$. In Fig. 7 we show the coupling coefficient $C_{11C}(0|0)$ as a function of $kd$. At each value of $kd$ the optimal radius $r_0$ has been chosen. For comparison  we show also the coupling coefficient $C_{11P}(0|0)$ as a function of $kd$ for the planar waveguide. This shows a decrease with increasing frequency of the fundamental. Clearly the circular waveguide geometry is to be preferred for most frequencies .

For large $N$ the Bragg factors $G$ in Eqs. (6.18) and (6.19) vary rapidly with frequency. It makes sense to compare the two geometries at their peak values for second harmonic generation. Hence we put the factor $F_N$ in Eq. (5.9) equal to $N^2$ in both cases. Then the Bragg factors differ only by the first factor in Eq. (5.7). Taking $w=a/2$ in both cases we get for the two Bragg factors at the peak value
\begin{equation}
\label{6.23}G_{N00C}=0.198\; N^2,\qquad G_{N00P}=0.377\;N^2.
\end{equation}
The prefactors  for the two geometries in the conversion coefficients (6.17) and (6.18) take similar values. The circular geometry may have an advantage over the planar geometry in the efficiency of input of the fundamental wave.

In the above numerical example we find for the circular waveguide with susceptibility $\chi_1=10^{-13}\;\mathrm{esu}$ the conversion factor in cgs units
\begin{equation}
\label{6.24}\eta^{MM}_{00C}=2.7\times 10^{-6}\;L^2I^M(\omega),\qquad (\mathrm{cgs})
\end{equation}
where $L=Na$ is the length of material with nonlinear susceptibility $\chi_1$. For an input power of the fundamental $I^M(\omega)=100\;\mathrm{mW}$ this becomes $\eta^{MM}_{00C}=2.7L^2$ with $L$ in cm. By definition the conversion factor is less than unity. This implies that for a length 1 cm we must take account of depletion.

\section{\label{VII}Depletion}

For a sufficiently long stretch of nonlinear susceptibility depletion of the fundamental must be taken into account. We consider a stretch with $M$ periods of nonlinear susceptibility $\vc{\chi}^{(2)}$, where $M$ is a large multiple of $N$. It is assumed that $N$ is sufficiently small that the preceding theory, with neglect of depletion, may be applied. In the experiment of Parameswaran et al. \cite{7} with a planar waveguide the number of periods is about $M=4000$. The period $a$ is assumed to be adjusted to the frequency of the fundamental by use of Eq. (5.10). Due to the Bragg factor conversion is then limited to the lowest mode with wavenumber $p_0(2\omega)$.

We assume first that the input laser is tuned to the peak value, so that the Bragg factor is proportional to $N^2$, as in Eq. (6.23).
On a large length scale the intensity of the fundamental $I^M_0(\omega;z)$ decreases with distance along the waveguide, and the intensity of the second harmonic $I^M_0(2\omega;z)$ increases. For brevity we denote $I_1(z)=I^M_0(\omega;z)$ and $I_2(z)=I^M_0(2\omega;z)$. We define corresponding slowly varying wave amplitudes $A_1(z)$ and $A_2(z)$ by \cite{2}
\begin{equation}
\label{7.1}I_1(z)=p_1A_1(z)^2I_1(0),\qquad I_2(z)=\frac{1}{2}p_2A_2(z)^2I_1(0),
\end{equation}
with the abbreviations
\begin{equation}
\label{7.2}p_1=p_0(\omega),\qquad p_2=p_0(2\omega).
\end{equation}
The amplitudes $A_1(z)$ and $A_2(z)$ are assumed to satisfy the mode-coupling equations
\begin{eqnarray}
\label{7.3}\frac{dA_1}{dz}&=&-\frac{\gamma}{p_1}A_1A_2,\nonumber\\
\frac{dA_2}{dz}&=&2\frac{\gamma}{p_2}A_1^2,
\end{eqnarray}
where the coefficient $\gamma$ can be evaluated from the preceding theory for the behavior at small $z$. The mode-coupling equations imply the conservation law
\begin{equation}
\label{7.4}I_1(z)+I_2(z)=I_1(0).
\end{equation}
The equations have the solution
\begin{equation}
\label{7.5}A_1(z)=\frac{1}{\sqrt{p_1}}\;\mathrm{sech}\;\kappa z,\qquad A_2(z)=\sqrt{\frac{2}{p_2}}\;\tanh\kappa z,
\end{equation}
with
\begin{equation}
\label{7.6}\kappa=\frac{\gamma}{p_1}\sqrt{\frac{2}{p_2}},
\end{equation}
so that correspondingly
\begin{equation}
\label{7.7}I_1(z)=I_1(0)\mathrm{sech}^2\kappa z,\qquad I_2(z)=I_1(0)\tanh^2\kappa z.
\end{equation}
Comparison with Eq. (6.11) yields
\begin{equation}
\label{7.8}\kappa^2=8\frac{a^2}{c}|e^{2\pi iw/a}-1|^2A_{00}I_1(0),
\end{equation}
with coupling factor $A_{00}$ given by Eq. (6.14). The decay length $1/\kappa$ decreases with increasing intensity of the incident laser light.

We note the identity
\begin{equation}
\label{7.9}\bigg(\frac{dA_1}{dz}\bigg)^2+\kappa^2A_1^2(p_1A_1^2-1)=0.
\end{equation}
This shows that the solution may be interpreted as the motion of a particle in a quartic potential at zero energy, or as the interface profile between two phases of a fluid \cite{17}. The identity is equivalent with the conservation law (7.4).

If the input laser is not tuned to the peak value we must use complex amplitudes $A_1(z),\;A_2(z)$ and generalize the mode-coupling equations to \cite{2},\cite{18}
\begin{eqnarray}
\label{7.10}\frac{dA_1}{dz}&=&-\frac{\gamma}{p_1}A_1^*A_2e^{i\Delta p\;z},\nonumber\\
\frac{dA_2}{dz}&=&2\frac{\gamma}{p_2}A_1^2e^{-i\Delta p\;z},
\end{eqnarray}
with phase mismatch
\begin{equation}
\label{7.11}\Delta p=p_2-2p_1-nK.
\end{equation}
With the normalization
\begin{equation}
\label{7.12}B_1=\sqrt{p_1}A_1,\qquad B_2=\frac{1}{2}\sqrt{p_2}A_2
\end{equation}
the mode-coupling equations can be expressed as
\begin{eqnarray}
\label{7.13}\frac{dB_1}{dz}&=&-\sigma B_1^*B_2e^{i\Delta p\;z},\nonumber\\
\frac{dB_2}{dz}&=&\frac{1}{2}\sigma B_1^2e^{-i\Delta p\;z},
\end{eqnarray}
with coefficient
\begin{equation}
\label{7.14}\sigma=\kappa\sqrt{2}.
\end{equation}
With the complex notation
\begin{equation}
\label{7.15}B_1=b_1e^{i\varphi_1},\qquad B_2=-ib_2e^{i\varphi_2},
\end{equation}
the mode-coupling equations (7.13) can be cast in the standard form with real variables \cite{2}
\begin{eqnarray}
\label{7.16}\frac{db_1}{dz}&=&-\sigma b_1b_2\sin\theta,\qquad\frac{d\varphi_1}{dz}=\sigma b_2\cos\theta,\nonumber\\
\frac{db_2}{dz}&=&\frac{1}{2}\sigma b_1^2\sin\theta,\qquad\frac{d\varphi_2}{dz}=\frac{1}{2}\sigma\frac{b_1^2}{b_2}\cos\theta,
\end{eqnarray}
with phase difference
\begin{equation}
\label{7.17}\theta=\Delta p\;z+\varphi_2-2\varphi_1.
\end{equation}
The mode-coupling equations imply the energy conservation law, which may be expressed as
\begin{equation}
\label{7.18}b_1^2(z)+2b_2^2(z)=1.
\end{equation}
With a final change of variables
\begin{equation}
\label{7.19}u=b_1,\qquad v=b_2\sqrt{2},\qquad \zeta=\kappa z
\end{equation}
the equations become
\begin{eqnarray}
\label{7.20}\frac{du}{d\zeta}&=&-uv\sin\theta,\nonumber\\
\frac{dv}{d\zeta}&=&u^2\sin\theta,\nonumber\\
\frac{d\theta}{d\zeta}&=&\Delta s+\cot\theta\frac{d}{d\zeta}\ln(u^2v),\qquad\Delta s=\frac{\Delta p}{\kappa}.
\end{eqnarray}
These equations have the two conservation laws
\begin{equation}
\label{7.21}u^2+v^2=1,\qquad u^2v\cos\theta+\frac{\Delta p}{2\kappa}\;v^2=\Gamma+\frac{\Delta p}{2\kappa}\;v_0^2,
\end{equation}
where the constant $\Gamma$ is determined by the initial values at $z=0$ according to $\Gamma=u_0v_0^2\cos\theta_0$. In our case $\Gamma=0$, since $v_0=0$. The equations can be solved in terms of the elliptic integral
\begin{equation}
\label{7.22}\zeta=\frac{1}{2}\int^{v^2}_0\frac{1}{\sqrt{x(1-x)^2-\frac{1}{4}\Delta s^2\;x^2}}\;dx.
\end{equation}
Explicitly
\begin{equation}
\label{7.23}\zeta=\sqrt{\gamma}\;F(\arcsin (v/\sqrt{\gamma})|\gamma^2),\qquad \gamma=8/(8+\Delta s^2+\sqrt{\Delta s^4+16\Delta s^2}),
\end{equation}
where $F(\varphi|\gamma^2)$ is the elliptic integral of the first kind \cite{19}. In particular $F(\varphi|1)=2\;\mathrm{arctanh}(\tan(\varphi/2))$ leads to $v=\tanh\zeta$ in agreement with Eq. (7.7).
The expression Eq. (7.23) can be inverted to
\begin{equation}
\label{7.24}v(\zeta)=\sqrt{\gamma}\;\mathrm{sn}(\zeta/\sqrt{\gamma},\gamma),
\end{equation}
with Jacobian elliptic function $\mathrm{sn}(\zeta/\sqrt{\gamma},\gamma)$. In Fig. 8 we plot the reduced second harmonic intensity $v^2$ for fixed length $\zeta=1$ as a function of the detuning parameter $\Delta s$. The plot shows a spectral line with sidewings. For fixed detuning parameter $\Delta s$ the second harmonic intensity varies periodically as a function of distance $\zeta$.

Note that in the present theory the mode-coupling equations are assumed to hold on a length scale much larger than the period of the grating. This is in contrast to the theory of Zhao et al. \cite{3}, who assume mode-coupling equations on the scale of the period. The present formulation allows understanding of the effect of depletion in the framework of the well-known theory developed by Armstrong et al. \cite{2} for second harmonic generation in anisotropic crystals.

Finally we note that it follows from Eq. (7.7) that for optimal phase matching the stretch of nonlinear susceptibility $L_{1/2}$ over which the second harmonic intensity $I_2$ is one half of the input fundamental intensity $I_1(0)$ is given by
\begin{equation}
\label{7.25}L_{1/2}=\frac{1}{\kappa}\;\mathrm{arctanh}\frac{1}{\sqrt 2}=\frac{0.881}{\kappa}.
\end{equation}
For the numerical example used in Eq. (6.24) we have $\kappa^2=2.7\times 10^{-6}I_1(0)$, so that
\begin{equation}
\label{7.26}L_{1/2}=\frac{5.36\times 10^{-13}}{\chi_1\sqrt{I_1(0)}}\;\mathrm{cm},
\end{equation}
with $\chi_1$ in esu and $I_1(0)$ in mW. The prefactor follows from $5.36=0.881/\sqrt{0.027}$. Thus the half-length varies inversely with the second order susceptibility and the square root of the input fundamental intensity. In Eq. (7.26) it is assumed that the radius of the tubelet $r_0$ has the optimal value shown in Fig. 4.

\section{\label{VIII}Discussion}

We have presented a model calculation of second harmonic generation in a circular cylindrical waveguide or optical fiber for a situation where the nonlinear material is isotropic and distributed in a radially symmetric manner. Though the geometry is not easy to realize experimentally, the calculation shows the essential features of the mechanism. A comparison with an earlier calculation for a planar waveguide \cite{1}, shows that the confinement in cylindrical geometry has distinct advantages. The work of Parameswaran et al. \cite{7} shows that second harmonic generation in cylindrical geometry can be realized experimentally and is quite effective.

The conversion efficiency is sufficiently high that depletion of the fundamental must be taken into account. We have shown that depletion can be described in terms of the mode-coupling formalism developed by Armstrong et al. \cite{2} for second harmonic generation in anisotropic crystals.

\newpage

\section*{Figure captions}

\subsection*{Fig. 1}
Axial cross-section of the waveguide and adsorbate structure. The figure should be rotated about the $z$ axis to get the three-dimensional picture.

\subsection*{Fig. 2}
Three-dimensional picture of the waveguide and adsorbate structure.

\subsection*{Fig. 3}
Plot of the wavenumber $p_0(k)$ of the lowest order guided wave (solid curve), of the wavenumber $p_1(k)$ of the first order guided wave (long dashes), of the wavenumber $p_2(k)$ of the second order guided wave (short dashes), as functions of $kd$ for values of the dielectric constant given in the text.

\subsection*{Fig. 4}
Plot of the coupling coefficient $C_{11}(0|0;r_0)$ for $kd=24$ as a function of $r_0/d$. The coefficient characterizes the efficiency of conversion, as given by Eqs. (6.12) and (6.14), of the lowest order guided wave with largest wavenumber $p_0(k)$ at the fundamental frequency $\omega=kc$ to the mode with largest wavenumber $p_0(2k)$ at the second harmonic frequency $2\omega=2kc$. The radius of the tubelet of nonlinear susceptibility is $r_0$, and $d$ is the radius of the core of the waveguide. The subscripts $11$ indicate the contribution which is quadratic in the component $\chi_1=\chi^{(2)}_{rrr}$ of the second order susceptibility tensor.

\subsection*{Fig. 5}
Plot of the coupling coefficient $C_{22}(0|0;r_0)$ for $kd=24$ as a function of $r_0/d$. The coefficient characterizes the efficiency of conversion, as given by Eqs. (6.12) and (6.14), of the lowest order guided wave with largest wavenumber $p_0(k)$ at the fundamental frequency $\omega=kc$ to the mode with largest wavenumber $p_0(2k)$ at the second harmonic frequency $2\omega=2kc$. The radius of the tubelet of nonlinear susceptibility is $r_0$, and $d$ is the radius of the core of the waveguide. The subscripts $22$ indicate the contribution which is quadratic in the component $\chi_2=\chi^{(2)}_{rzz}$ of the second order susceptibility tensor.

\subsection*{Fig. 6}
Plot of the coupling coefficient $C_{33}(0|0;r_0)$ for $kd=24$ as a function of $r_0/d$. The coefficient characterizes the efficiency of conversion, as given by Eqs. (6.12) and (6.14), of the lowest order guided wave with largest wavenumber $p_0(k)$ at the fundamental frequency $\omega=kc$ to the mode with largest wavenumber $p_0(2k)$ at the second harmonic frequency $2\omega=2kc$. The radius of the tubelet of nonlinear susceptibility is $r_0$, and $d$ is the radius of the core of the waveguide. The subscripts $33$ indicate the contribution which is quadratic in the component $\chi_1=\chi^{(2)}_{zzr}$ of the second order susceptibility tensor.

\subsection*{Fig. 7}
Plot of the coupling coefficient $C_{11}(0|0)$ for the circular waveguide as a function of $kd$ (solid curve). For each value of $kd$ the optimal radius $r_0$ has been chosen. For comparison we also plot the coupling coefficient $C_{11}(0|0)$ for the planar waveguide as a function of $kd$ (dashed curve). The notation is explained in the caption to Fig. 4.

\subsection*{Fig. 8}
Plot of the reduced second harmonic intensity $v^2$ at fixed length $L=1/\kappa$ as a function of the reduced detuning parameter $\Delta s=\Delta p/\kappa$, where $\Delta p$ is the phase mismatch. The notation is explained in Sec. 7.

\newpage
\setlength{\unitlength}{1cm}
\begin{figure}
 \includegraphics{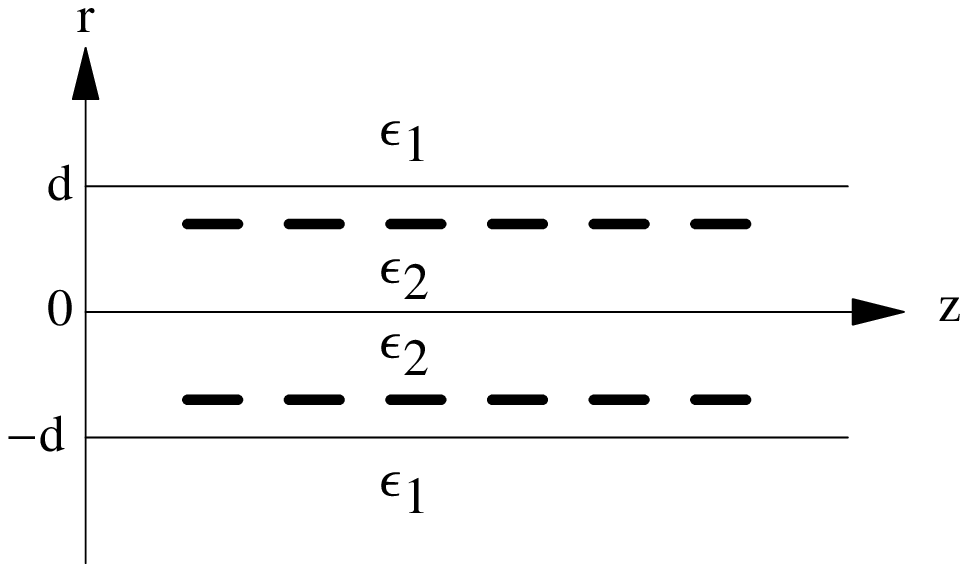}

  \caption{}
\end{figure}
\newpage
\clearpage
\newpage
\setlength{\unitlength}{1cm}
\begin{figure}
 \includegraphics{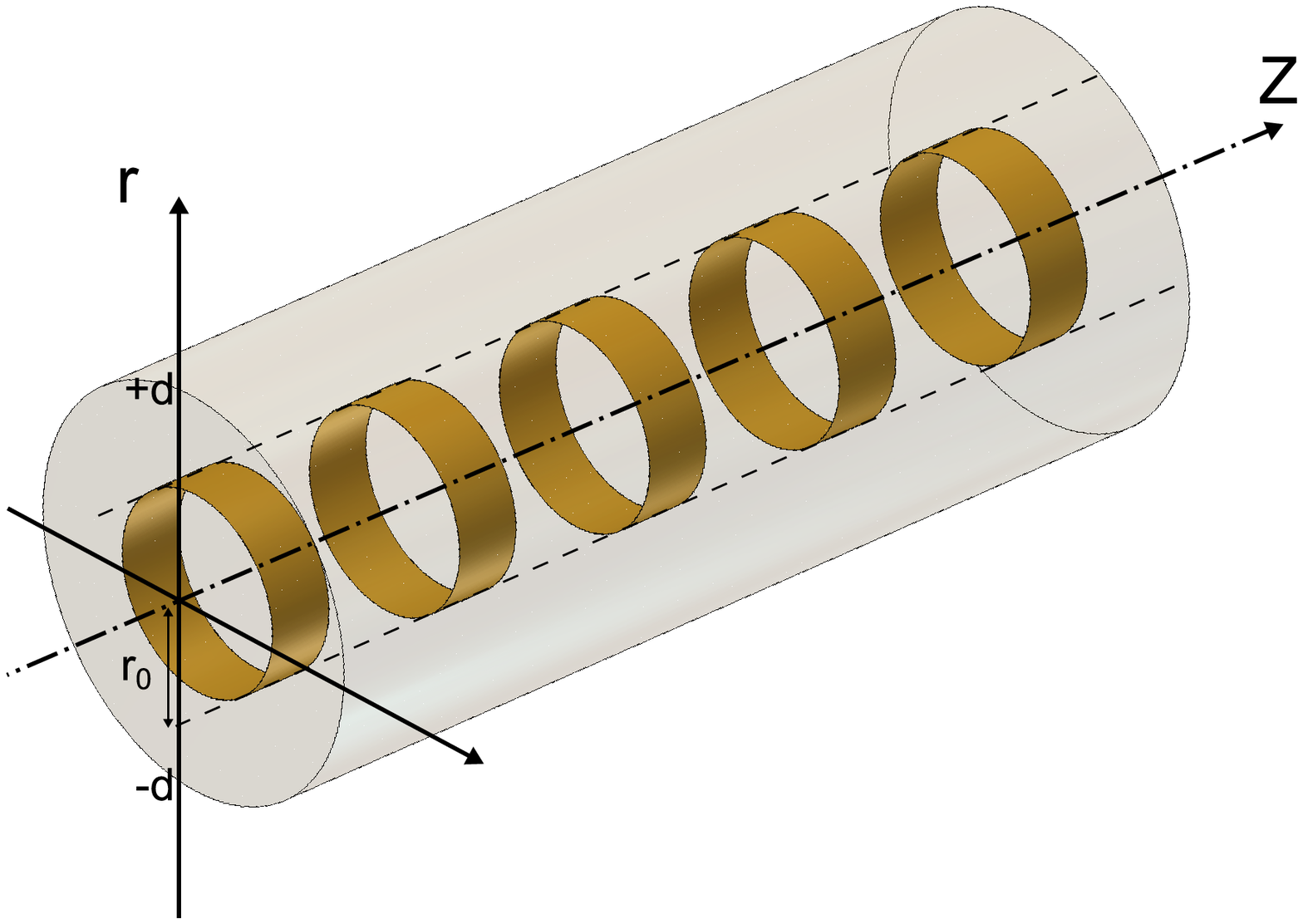}

  \caption{}
\end{figure}
\newpage
\clearpage
\newpage
\setlength{\unitlength}{1cm}
\begin{figure}
 \includegraphics{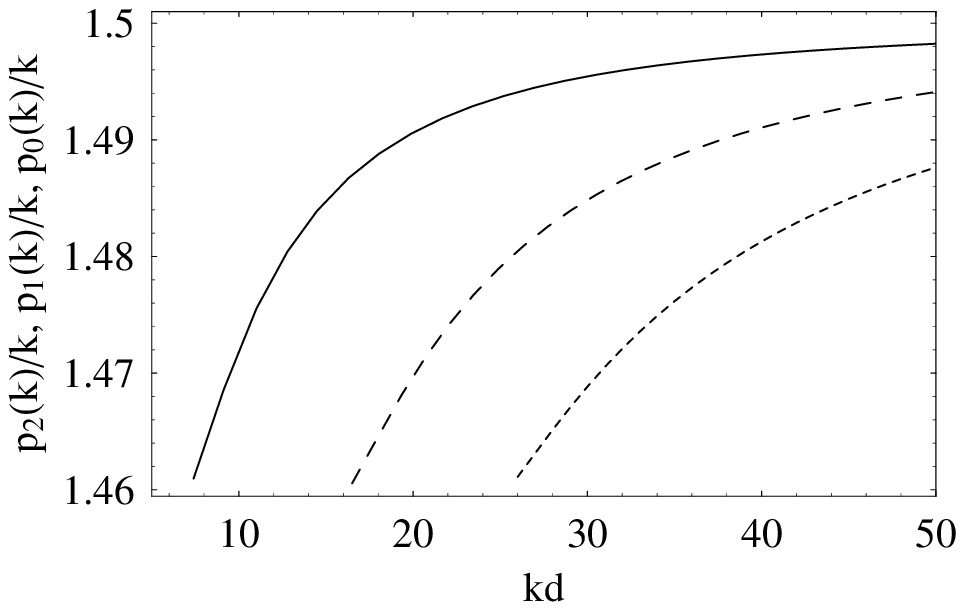}

  \caption{}
\end{figure}
\newpage
\clearpage
\newpage
\setlength{\unitlength}{1cm}
\begin{figure}
 \includegraphics{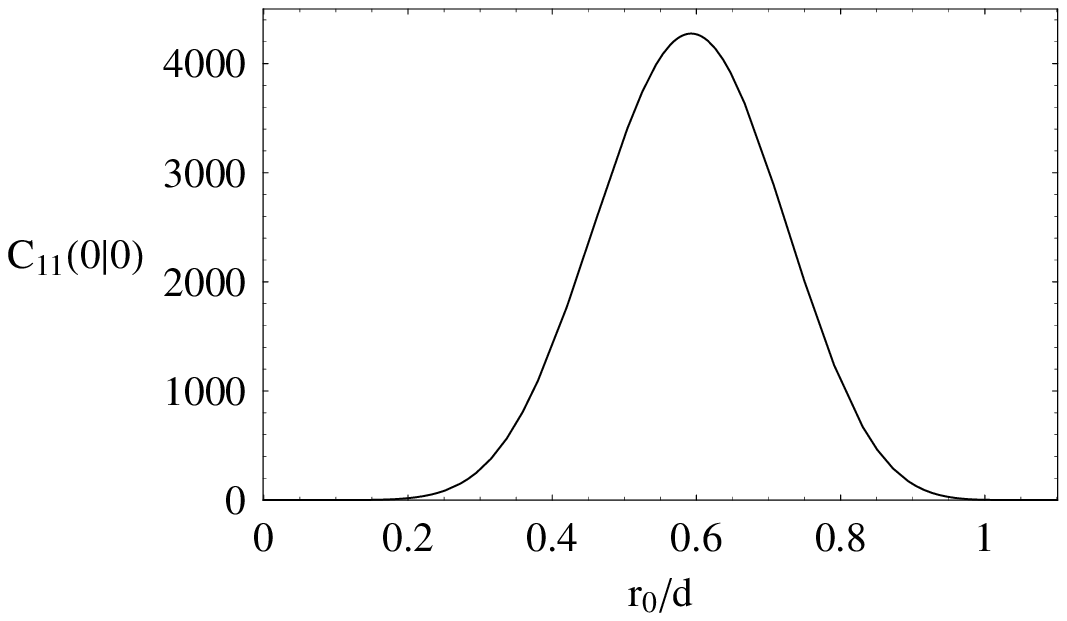}

  \caption{}
\end{figure}
\newpage
\clearpage
\newpage
\setlength{\unitlength}{1cm}
\begin{figure}
 \includegraphics{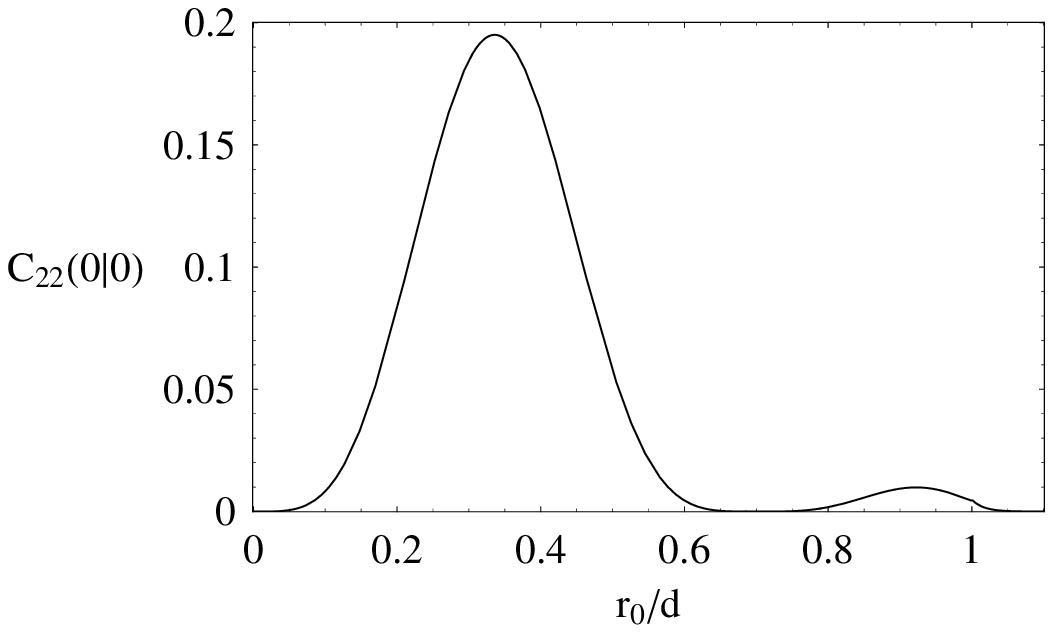}

  \caption{}
\end{figure}
\newpage
\clearpage
\newpage
\setlength{\unitlength}{1cm}
\begin{figure}
 \includegraphics{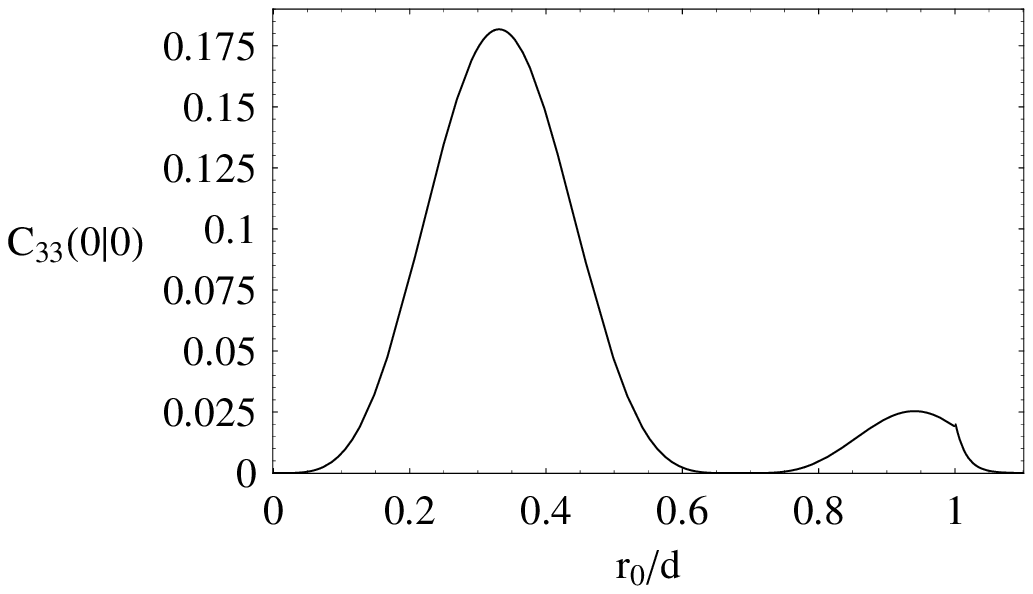}

  \caption{}
\end{figure}
\newpage
\clearpage
\newpage
\setlength{\unitlength}{1cm}
\begin{figure}
 \includegraphics{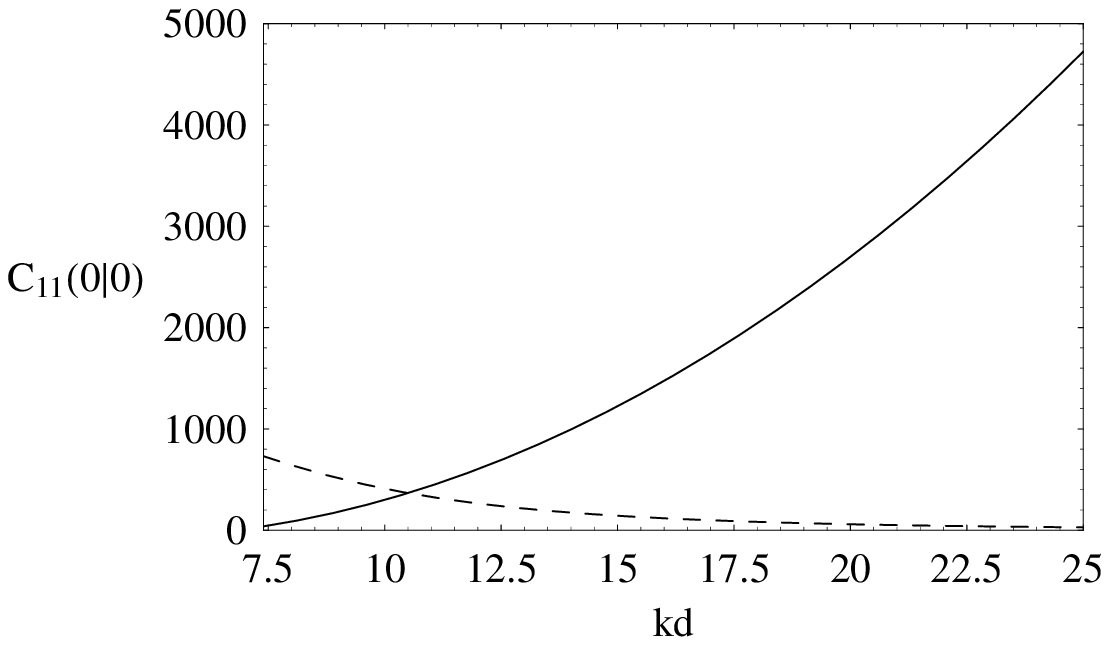}

  \caption{}
\end{figure}
\newpage
\clearpage
\newpage
\setlength{\unitlength}{1cm}
\begin{figure}
 \includegraphics{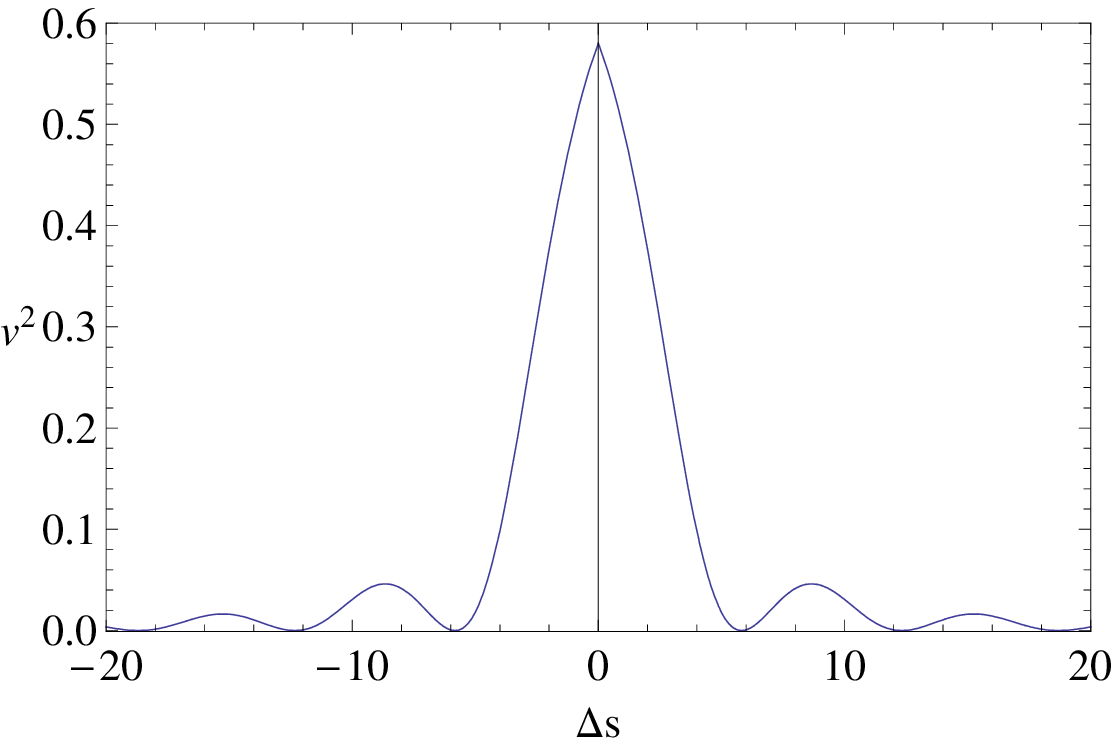}

  \caption{}
\end{figure}
\newpage

\end{document}